\documentclass[9pt,shortpaper,twoside,web]{ieeecolor}
\usepackage{generic}
\usepackage{cite}
\usepackage{amsmath,amssymb,amsfonts}
\usepackage{graphicx}
\usepackage{textcomp}

\usepackage{epsfig} 
\usepackage{times} 
\usepackage{url}
\usepackage{algorithm,algorithmic}

\usepackage{color}

\usepackage{theorem}
\theorembodyfont{\rmfamily}

\newtheorem{lem}{Lemma}
\newtheorem{defin}{Definition}
\newtheorem{theorem}{Theorem}
\newtheorem{prop}{Proposition}
\newtheorem{assump}{Assumption}

\renewcommand{\QED}{\QEDopen}

\usepackage[caption=false,font=footnotesize]{subfig}

\newcommand{\dg}{^\dagger}
\newcommand{\ol}[1]{\overline{#1}}
\newcommand{\hs}{& \hspace{-3mm}}
\newcommand{\tl}[1]{\tilde{#1}}

\def\BibTeX{{\rm B\kern-.05em{\sc i\kern-.025em b}\kern-.08em
    T\kern-.1667em\lower.7ex\hbox{E}\kern-.125emX}}
\begin{document}
\title{Parameterization of All Output-Rectifying Retrofit Controllers}
\author{Hampei Sasahara, \IEEEmembership{Member,~IEEE},
Takayuki Ishizaki, \IEEEmembership{Member,~IEEE},\\
and Jun-ichi Imura, \IEEEmembership{Senior Member,~IEEE}
\thanks{Manuscript received xxx xx, 20xx; revised xxx xx, 20xx.}
\thanks{H.~Sasahara is with the Division of Decision and Control Systems, KTH Royal Institute of Technology, Stockholm, SE-100 44 Sweden e-mail: hampei@kth.se.}
\thanks{T.~Ishizaki and J.~Imura are with the Graduate School
of Engineering, Tokyo Institute of Technology,
Tokyo, 152-8552 Japan e-mail: \{ishizaki, imura\}@sc.e.titech.ac.jp.}
\thanks{This work was supported by JST MIRAI Grant Number 18077648, Japan, and TEPCO Memorial Foundation.}
\thanks{\copyright 2021 IEEE.  Personal use of this material is permitted.  Permission from IEEE must be obtained for all other uses, in any current or future media, including reprinting/republishing this material for advertising or promotional purposes, creating new collective works, for resale or redistribution to servers or lists, or reuse of any copyrighted component of this work in other works.}
}

\maketitle

\begin{abstract}
This study investigates a parameterization of all output-rectifying retrofit controllers for distributed design of a structured controller.
It has been discovered that all retrofit controllers can be characterized as a constrained Youla parameterization, which is difficult to solve analytically.
For synthesis, a tractable and insightful class of retrofit controllers, referred to as output-rectifying retrofit controllers, has been introduced.
An unconstrained parameterization of all output-rectifying retrofit controllers can be derived under a technical assumption on measurability of the interaction signal.
The aim of this note is to reveal the structure of all output-rectifying retrofit controllers in the general output-feedback case without interaction measurement.
It is found out that the existing developments can be generalized based on system inversion.
The result leads to the conclusion that output-rectifying retrofit controllers can readily be designed even in the general case.
\end{abstract}

\begin{IEEEkeywords}
Distributed design, large-scale systems, network systems, retrofit control, Youla parameterization.
\end{IEEEkeywords}
\section{Introduction}

This note addresses \emph{distributed design} of subcontrollers constituting a structured controller in a large-scale network system.
The traditional controller design methods, e.g., decentralized and distributed control~\cite{Sandell1978Survey,Bakule2008Decentralized,Siljak2011Decentralized}, are built on the premise that there exists a unique controller designer, while there are often multiple independent controller designers in practical network systems.
While integrated controller design by a unique designer is referred to as centralized design, independent design of subcontrollers by multiple designers is referred to as distributed design~\cite{Langbort10}.
The primary difficulty of distributed design is that, from the perspective of a single controller designer, even if the entire model information is provided at some time instant, the dynamics possibly varies depending on other controller designers' actions.

\emph{Retrofit control}~\cite{Ishizaki2018Retrofit,Ishizaki2019Modularity,Ishizaki2019Retrofit} is a promising approach for distributed design.
In its framework, the network system to be controlled is regarded as an interconnected system composed of the subsystem of interest and its environment composed of the other unknown subsystems with their interaction.
Retrofit controllers are defined as the controllers that can guarantee internal stability of the network system for any possible environment as long as the network system to be controlled, itself, is initially stable.
By designing a retrofit controller as an add-on subcontroller, each subcontroller designer can introduce her own control policy independently of the others.
It has been discovered that all retrofit controllers can be characterized through the Youla parameterization with a linear constraint on the Youla parameter~\cite{Ishizaki2019Modularity}.
Unfortunately, because the constrained parameterization is difficult to handle analytically, synthesis of the most general retrofit controller cannot be performed in a straightforward manner.

To resolve this issue, a particular class of retrofit controllers, referred to as output-rectifying retrofit controllers, has been introduced.
A systematic design method has been proposed in~\cite{Ishizaki2018Retrofit} where a technical assumption on measurability of signals that contain adequate information on environment's behavior is made for simplifying the arguments.
Specifically, as the most basic case, the situation where the inflowing interaction signal from the environment is measurable has been addressed.
The proposed approach is further extended to the state-feedback case~\cite{Ishizaki2018Retrofit}.
Moreover, it has been found out in~\cite{Ishizaki2019Modularity} that the proposed structure is also necessary for output-rectifying retrofit controllers when the interaction signal is measurable.
This finding leads to a parameterization with which the output-rectifying retrofit controller design problem can be reduced to a standard controller design problem.
However, a parameterization of output-rectifying retrofit controllers in the general output-feedback case still remains an open problem.

The objective of this note is to provide a systematic design method by solving the problem theoretically.
The difficulty of the generalization is that all solutions to linear equations over the ring of real, rational, and stable transfer matrices are needed to be characterized.
The fundamental idea is to build appropriate bases that span the solution space based on system inversion.
It is found out that an unconstrained parameterization of all output-rectifying retrofit controllers can be obtained even in the general case.
A preliminary version of this work was presented in the publication~\cite{Sasahara2018Parameterization}, where its analysis is limited only to the state-feedback case.

This note is organized as follows.
In Sec.~\ref{sec:review}, we review the exiting retrofit control framework and pose the problem of interest.
Sec.~\ref{sec:of} generalizes the existing result to the output-feedback case without interaction measurement.
In Sec.~\ref{sec:num}, we numerically verify the effectiveness of the obtained results.
Sec.~\ref{sec:conc} draws the conclusion.

{\it Notation:}
We denote the set of the real numbers by $\mathbb{R}$,
the set of the $n \times m$ real matrices by $\mathbb{R}^{n \times m}$,
the identity matrix by $I$,
a pseudo inverse of a matrix $M$ by $M\dg$,
the matrix where matrices $M_i$ for $i=1,\ldots,m$ are concatenated vertically by ${\rm col}(M_i)_{i=1}^m$,
the block-diagonal matrix whose diagonal blocks are composed of $M_i$ for $i=1,\ldots,m$ by ${\rm diag}(M_i)_{i=1}^m$,
the set of real and rational $n \times m$ transfer matrices by $\mathcal{R}^{n \times m}$,
the set of proper transfer matrices in $\mathcal{R}^{n \times m}$ by $\mathcal{RP}^{n \times m}$,
and the set of stable transfer matrices in $\mathcal{RP}^{n\times m}$ by $\mathcal{RH}^{n \times m}_{\infty}$.
When the dimensions of the spaces are clear from the context, the superscripts are omitted.
A transfer matrix $K$ is said to be a stabilizing controller for $G$ if the feedback system of $G$ and $K$ is internally stable, i.e., the four transfer matrices
$(I-KG)^{-1}K, (I-KG)^{-1}, (I-GK)G^{-1},$ and $(I-GK)^{-1}$
belong to $\mathcal{RH}_{\infty}$~\cite[Chap.~5]{Zhou1996Robust}.
The set of all stabilizing controllers in $\mathcal{RP}$ for $G$ is denoted by $\mathcal{S}(G)$.
Note that if $G$ is stable, the internal stability is equivalent to stability of $(I-KG)^{-1}K$, which is often denoted by $Q$ and referred to as the Youla parameter~\cite[Chap.~12]{Zhou1996Robust}.

\section{Brief Review of Retrofit Control}
\label{sec:review}

In this section, we first review the retrofit control based on the formulation in~\cite{Ishizaki2019Modularity}.
Further, we pose the problems treated in this paper.

\subsection{Retrofit Control Framework: Modular Subcontroller Design}
\label{subsec:retro_framework}

\begin{figure}[t]
\centering
\subfloat[][A network system composed of $N$ subsystems with an interaction and a decentralized controller.]{\includegraphics[width=.47\linewidth]{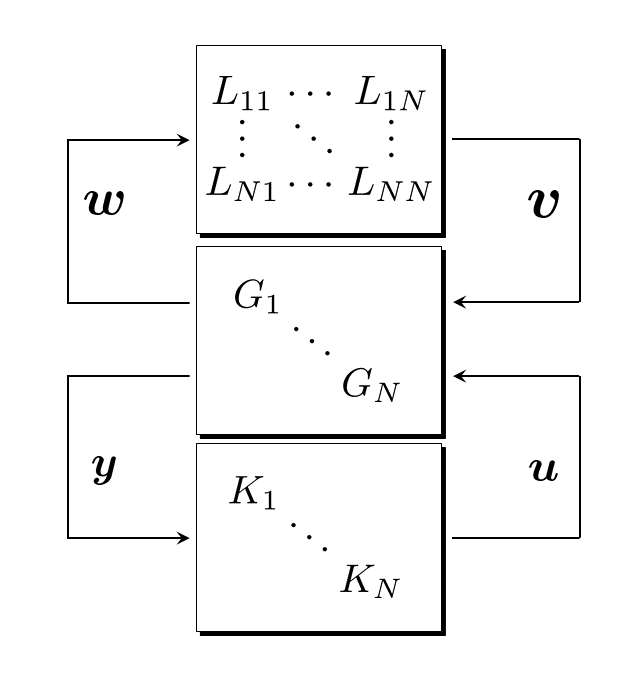}\label{subfig:a_sys}}
\quad
\subfloat[][Example of modular subcontroller design with two subsystems from the viewpoint of the first subcontroller designer.]{\includegraphics[width=.47\linewidth]{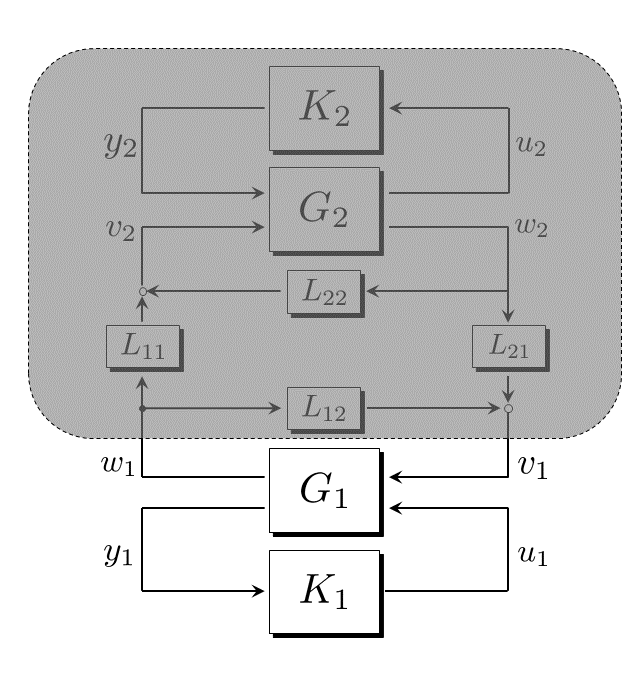}\label{subfig:a_sys_ex}}
\caption{The systems considered in the retrofit control framework.}
\label{fig:sys_retrofit}
\end{figure}

The aim of retrofit control is to enable modular design of subcontrollers for network systems, i.e., parallel synthesis via multiple independent subcontroller designers to handle complexity of large-scale system design.
In the retrofit control framework, we consider a network system, whose model is depicted in Fig.~\ref{subfig:a_sys}, where $N$ subsystems $G_1,\ldots,G_N$ are interconnected through $L_{ij}$ for $i,j=1,\ldots,N$ and governed by a decentralized controller consisting of $K_1,\ldots,K_N$.
The signals $\boldsymbol{v,w,u,y}$ represent the stacked vectors of the inflowing interaction signals, the outflowing interaction signals, the control inputs, and the measurement signals, respectively.
It is supposed that there are $N$ subcontroller designers each of whom is responsible for designing her corresponding subcontroller only with the model information on her own subsystem.
As an example, a schematic diagram of the network system with two subsystems from the viewpoint of the first subcontroller designer is illustrated in Fig.~\ref{subfig:a_sys_ex} where only the model information on $G_1$ is available to the designer of $K_1$.

Ensuring stability of the closed-loop system under modular design can be a difficult problem, since the overall dynamics depends on all designers' control policies.
Retrofit control has been proposed for resolving this issue.
In the next subsection, we give the formal definition of retrofit controllers.

\subsection{Definition of Retrofit Controllers}
\begin{figure}[t]
\centering
\includegraphics[width = .95\linewidth]{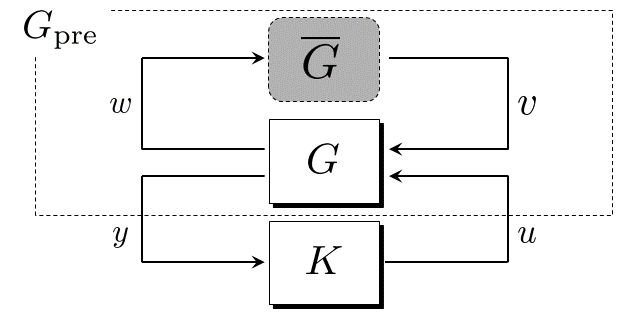}
\caption{The interconnected system from the perspective of each subcontroller designer in the retrofit control framework.
The block $K$ represents the retrofit controller to be designed, and the blocks $G$ and $\ol{G}$ represent the subsystem of interest and its unknown environment, respectively.
The preexisting system $G_{\rm pre}$ represents the original interconnected system composed of $G$ and $\ol{G}$ without $K$.
}
\label{fig:pre_sys}
\end{figure}

In this note, we consider an interconnected system in Fig.~\ref{fig:pre_sys} where
\begin{equation}\label{eq:sys_tf}
 \left[
  \begin{array}{c}
  w\\
  y
  \end{array}
 \right] = \underbrace{\left[
 \begin{array}{cc}
 G_{wv} \hs G_{wu}\\
 G_{yv} \hs G_{yu}
 \end{array}
 \right]}_{G}\left[
 \begin{array}{c}
 v\\
 u
 \end{array}
 \right]
\end{equation}
is referred to as a \emph{subsystem of interest} for retrofit control,
and $v=\ol{G}w$
is referred to as its \emph{environment}.
The interconnected system from $u$ to $y$ is given by
\[
 \begin{array}{rl}
 G_{\rm pre} := \hs G_{yu} + G_{yv}\ol{G}(I-G_{wv}\ol{G})^{-1}G_{wu},
 \end{array}
\]
to which we refer as the \emph{preexisting system}.
In the system representation, $v,w$ denote the inflowing and outflowing interaction signals and
$u,y$ denote the control input and the measurement output.
We describe a state-space representation of the subsystem of interest~\eqref{eq:sys_tf} as
\begin{equation}\label{eq:Gss}
 G: \left\{
 \begin{array}{cl}
 \dot{x} \hs = Ax+Lv+Bu\\
 w \hs = {\it \Gamma}x\\
 y \hs = Cx
 \end{array}
 \right.
\end{equation}
where $x$ is the state of $G$.
The dynamical controller to be designed is given by $K$ generating the control input according to $u=Ky$.
It should be noted that, although an exogenous input and an evaluation output are not considered because this note focuses just on stability analysis, our framework can also discuss control performance~\cite{Ishizaki2019Modularity}.

In this description, $G$ and $K$ represent a single designer's subsystem and her subcontroller, respectively, and $\ol{G}$ represents the other subsystems, the other subcontrollers, and their interaction.
For example, for the system in Fig.~\ref{subfig:a_sys_ex}, $G$ and $K$ are given as $G_1$ and $K_1$, respectively, and $\ol{G}$ is given as the other components.

As stated in Sec.~\ref{subsec:retro_framework}, we suppose that the preexisting system is initially stable or has been stabilized by a controller inside the environment $\ol{G}$ aside from the controller $K$ as in~\cite{Ishizaki2019Modularity}.
Under this assumption, in order to reflect the obscurity of the model information on $\ol{G}$, we introduce the set of admissible environments as
\[
 \ol{\mathcal{G}} := \{\ol{G} \in \mathcal{RP}: G_{\rm pre}\ {\rm is\ internally\ stable}\}.
\]

The purpose of retrofit control is to design the controller $K$ to improve a control performance without losing internal stability of the whole network system.
Accordingly, we define \emph{retrofit controllers} as follows.
\begin{defin}
The controller $K$ is said to be a \emph{retrofit controller} if the resultant control system is internally stable for any environment $\ol{G} \in \ol{\mathcal{G}}$.
\end{defin}
Retrofit control enables distributed design of subcontrollers for a network system.
By designing a retrofit controller as an add-on controller, each subcontroller designer can introduce her own controller independently of the others.

\subsection{Characterization of All Retrofit Controllers}
\label{subsec:chara}

\begin{figure}[t]
\centering
\subfloat[][Closed-loop block diagram.]{\includegraphics[width=.47\linewidth]{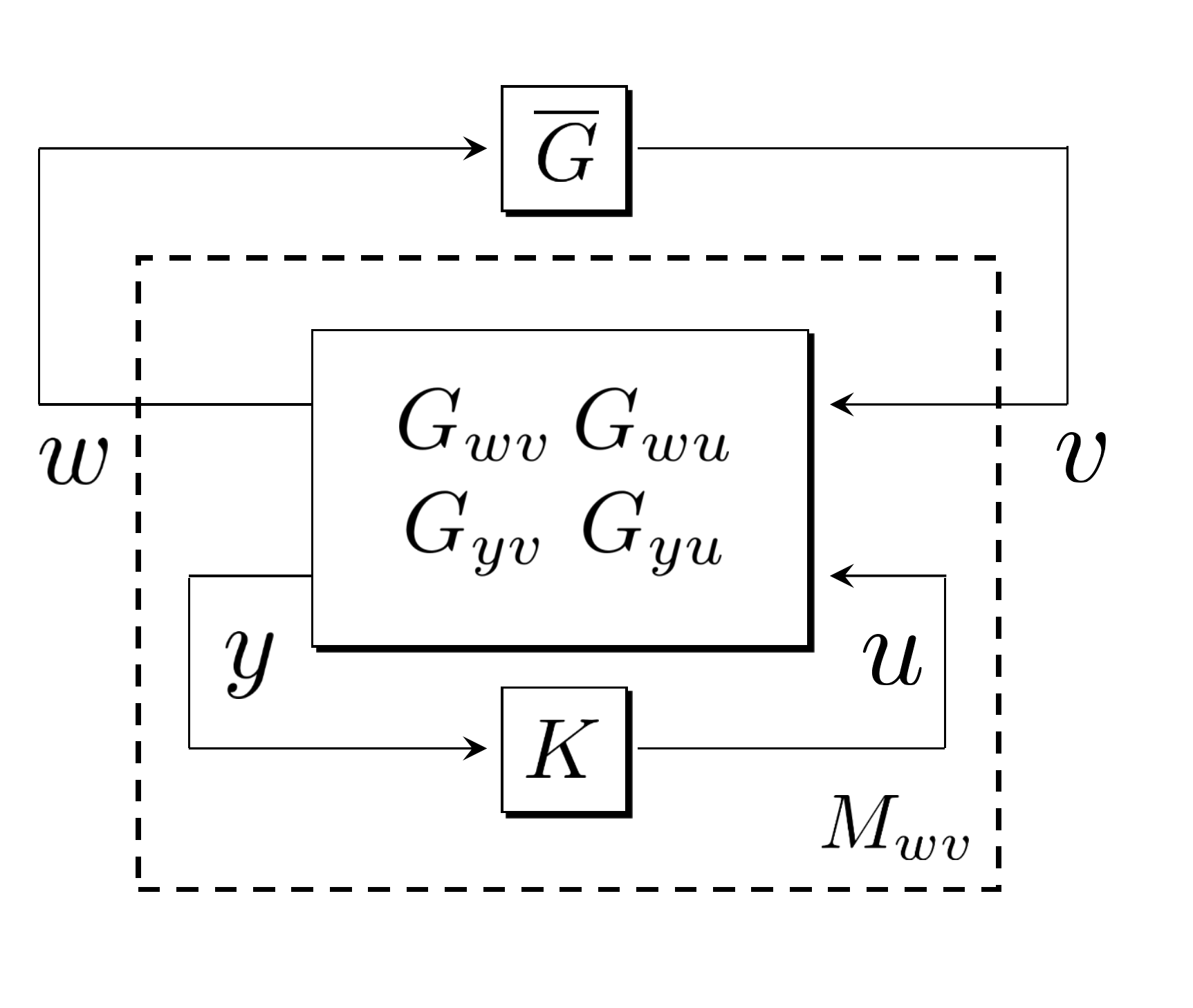}\label{subfig:M_stable}}\quad\subfloat[][Closed-loop block diagram with the decomposition $\ol{G}=\ol{G}_0+\Delta\ol{G}$.]{\includegraphics[width=.47\linewidth]{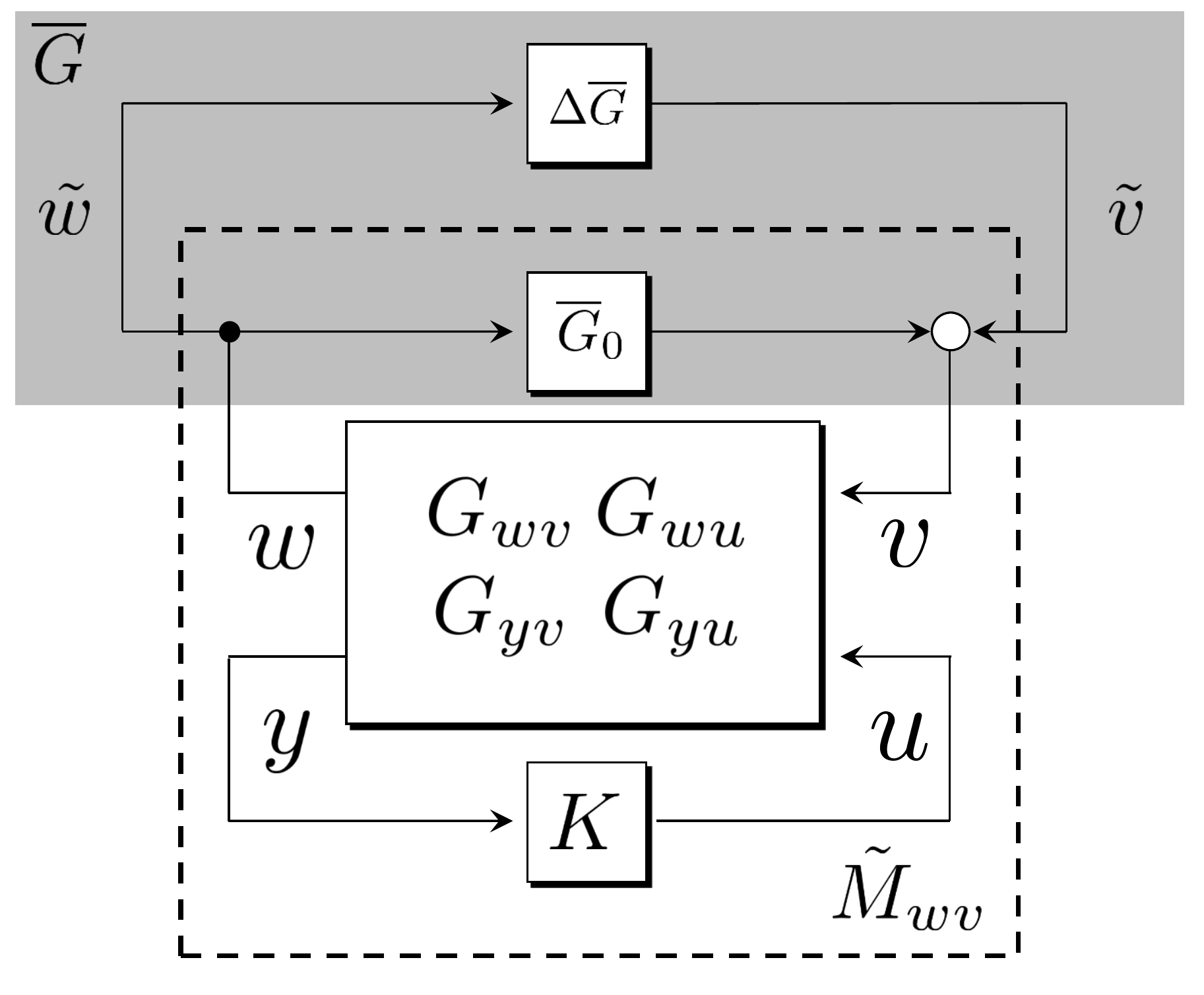}\label{subfig:M_unstable}}
\caption[]{Block diagrams of the closed-loop system.}
\label{fig:cl_sys}
\end{figure}

The basic idea of designing a retrofit controller is to preserve the internal stability of the preexisting system by maintaining the dynamical relationship between the interaction signals to be invariant.
As a preliminary step, we here pay attention only to stable subsystems.
The following assumption is made.
\begin{assump}\label{assum:sta}
 The subsystem $G$ is stable, i.e., $G\in \mathcal{RH}_{\infty}$.
\end{assump}
Then the aforementioned idea is mathematically described by
\begin{equation}\label{eq:invariance}
 M_{wv}=G_{wv}
\end{equation}
where $M_{wv}:=G_{wv}+G_{wu}(I-KG_{yu})^{-1}KG_{yv}$ denotes the closed-loop transfer matrix from $v$ to $w$ in Fig.~\ref{subfig:M_stable}.
The condition~\eqref{eq:invariance} is equivalent to
\begin{equation}\label{eq:GQG}
 G_{wu}QG_{yv}=0
\end{equation}
where $Q:=(I-KG_{yu})^{-1}K$ is the Youla parameter of $K$ for $G_{yu}$.
The first existing result claims that this condition~\eqref{eq:invariance} or its alternative~\eqref{eq:GQG} are necessary and sufficient conditions for retrofit control~\cite{Ishizaki2019Modularity}.
\begin{prop}\label{prop:GQG}
Let Assumption~\ref{assum:sta} hold.
Then $K$ is a retrofit controller if and only if the Youla parameter $Q \in \mathcal{RH}_{\infty}$ satisfies~\eqref{eq:GQG}.
\end{prop}

This argument can be extended to the general case without Assumption~\ref{assum:sta}, i.e., $G$ is possibly unstable.
Consider decomposing the environment as  $\ol{G}=\ol{G}_0+\Delta\ol{G}$ where $\ol{G}_0$ is a certain transfer matrix such that the feedback system composed of $G$ and $\ol{G}_0$ is internally stable.
By regarding the subsystem $G$ with $\ol{G}_0$ as a modified subsystem of interest and $\Delta \ol{G}$ as its environment, we obtain a replacement of the condition~\eqref{eq:invariance} as $\tl{M}_{wv}=\tl{G}_{wv},$
where $\tl{M}_{wv}$ and $\tl{G}_{wv}$ are defined to be the closed-loop and open-loop transfer matrices from $\tl{v}$ to $\tl{w}$ in Fig.~\ref{subfig:M_unstable}, respectively.
Indeed, this condition is necessary and sufficient for retrofit control with respect to possibly unstable subsystems~\cite{Ishizaki2019Modularity}.
As conducted above, we can systematically transform an unstable subsystem into a stable one, and hence we let Assumption~\ref{assum:sta} hold throughout this paper to avoid notational burden.

\subsection{Retrofit Controller Synthesis: Output-Rectifying Retrofit Controllers}

For retrofit controller synthesis, it suffices to find an appropriate Youla parameter $Q \in \mathcal{RH}_{\infty}$ that satisfies the constraint~\eqref{eq:GQG} under a desired performance criterion.
However, this constraint is difficult to handle analytically.
Furthermore, even if we obtain a solution numerically, the internal structure of the resulting controller is unclear.
To design an insightful retrofit controller, a tractable class of retrofit controllers has been introduced~\cite{Ishizaki2019Modularity}.
\begin{defin}\label{def:out}
The controller $K$ is said to be an \emph{output-rectifying retrofit controller} if
\begin{equation}\label{eq:QG}
 QG_{yv}=0
\end{equation}
where $Q \in \mathcal{RH}_{\infty}$ denotes the Youla parameter of $K$ for $G_{yu}$.
\end{defin}
Obviously, if $K$ satisfies these conditions, then $K$ is a retrofit controller.

\begin{figure}[t]
\centering
\includegraphics[width = .95\linewidth]{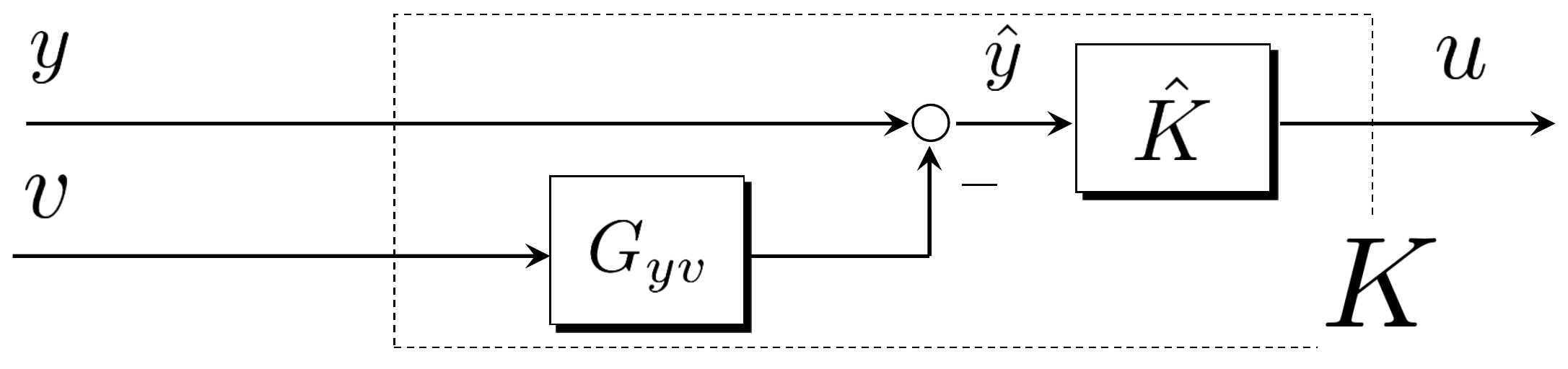}
\caption{The internal structure of all output-rectifying retrofit controllers where the internal controller $\hat{K}$ is a design parameter given as a stabilizing controller for $G_{yu}$.
}
\label{fig:retro_str}
\end{figure}

This class is tractable in the sense that output-rectifying retrofit controllers can be designed through a standard controller design method when the inflowing interaction signal is measurable.
The fundamental idea is to rectify the input signal injected into an internal controller $\hat{K}$ as
\begin{equation}\label{eq:rect_ex}
 \hat{y}:=y-G_{yv}v.
\end{equation}
In this architecture, the measurement output is rectified to be $\hat{y}$ so as to remove the effect of $v$ to $y$ through the \emph{rectifier}
\begin{equation}\label{eq:ex_rect}
 R:= [I\ -G_{yv}].
\end{equation}
Indeed, this structure characterizes all output-rectifying retrofit controllers with interaction measurement~\cite{Ishizaki2019Modularity}.
\begin{prop}\label{prop:out_str}
Let Assumption~\ref{assum:sta} hold.
Further, assume that the inflowing interaction signal $v$ is measurable in addition to $y$.
Then $K$ is an output-rectifying retrofit controller if and only if there exists $\hat{K} \in \mathcal{S}(G_{yu})$ such that $K=\hat{K}R$ with $R$ in~\eqref{eq:ex_rect}.
\end{prop}
Proposition~\ref{prop:out_str} implies that an output-rectifying retrofit controller can be designed by imposing the structure depicted by Fig.~\ref{fig:retro_str} into the controller to be designed with a parameter $\hat{K}$ that stabilizes $G_{yu}$.
It is also implied that this is the only procedure to design an output-rectifying retrofit controller.
This proposition also explains the reason for terming as ``output-rectifying'' in Definition~\ref{def:out}.
Although input-rectifying retrofit controllers can be defined as a dual notion~\cite{Ishizaki2019Modularity},
we consider only output-rectifying retrofit controllers in this paper.

\subsection{Objective of This Note}

\begin{figure}[t]
\centering
\includegraphics[width = .95\linewidth]{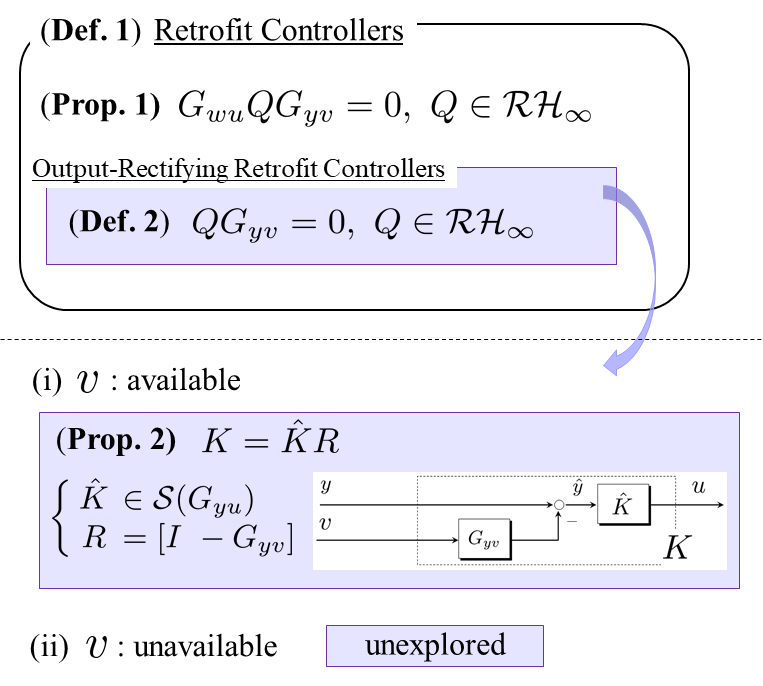}
\caption{
Venn diagrams of retrofit controllers based on the existing results in~\cite{Ishizaki2019Modularity}.
}
\label{fig:Venn}
\end{figure}

The reviewed existing results on output-rectifying retrofit controllers are summarized in Fig.~\ref{fig:Venn}, where Venn diagrams of retrofit controllers are described.
When the interaction signal can be measured, a complete parameterization of all output-rectifying retrofit controllers can be obtained as described in~(i).
On the other hand, the parameterization without interaction measurement, described in~(ii), has not been explored so far.
The objective of this paper is to complement the results by revealing the internal structure of output-rectifying retrofit controllers.
In particular, it is shown that the structure proposed in the existing work provides a parameterization in the general case.

\section{Parameterization of Output-Rectifying Retrofit Controllers without Interaction Measurement}
\label{sec:of}

\subsection{Basic Idea: Inverse System}
In this section, we generalize Proposition~\ref{prop:out_str} to the output-feedback case without interaction measurement.
The idea of the extension is to reproduce $v$ from $y$ through an inverse system.

For simplifying discussion, the following assumption is made.
\begin{assump}\label{assum:inv}
The transfer matrix $G_{yv}$ is not right-invertbile in $\mathcal{R}$,
and in addition, $G_{yv}$ is left-invertible in $\mathcal{R}$.
\end{assump}
Assumption~\ref{assum:inv} can be made without loss of generality.
The first assumption is made just for excluding the trivial case.
If $G_{yv}$ is right-invertible, then the condition~\eqref{eq:QG} is equivalent to $Q=0$, under which only the trivial controller $K=0$ exists.
On the other hand, the second assumption is made for simplifying the technical discussion.
If $G_{yv}$ is not left-invertible, which means that the interaction signal has redundancy, then the null space of $G_{yv}$ in $\mathcal{R}$ becomes a nonzero-dimensional subspace.
Thus there exists a decomposition $G_{yv}=G'_{yv}G_0$ with a left-invertible transfer matrix $G'_{yv}$ and a right-invertible transfer matrix $G_0$.
Then the condition~\eqref{eq:QG} is equivalent to $QG'_{yv}=0$, and it suffices to consider $G'_{yv}$ instead of $G_{yv}$.
Note that the invertibility is taken in $\mathcal{R}$, i.e., the left inverse can be improper.
Although it seems to result in improper controllers, we actually derive a parameterization of \emph{proper} controllers with the inverse system.

\begin{figure}[t]
\centering
\includegraphics[width = .95\linewidth]{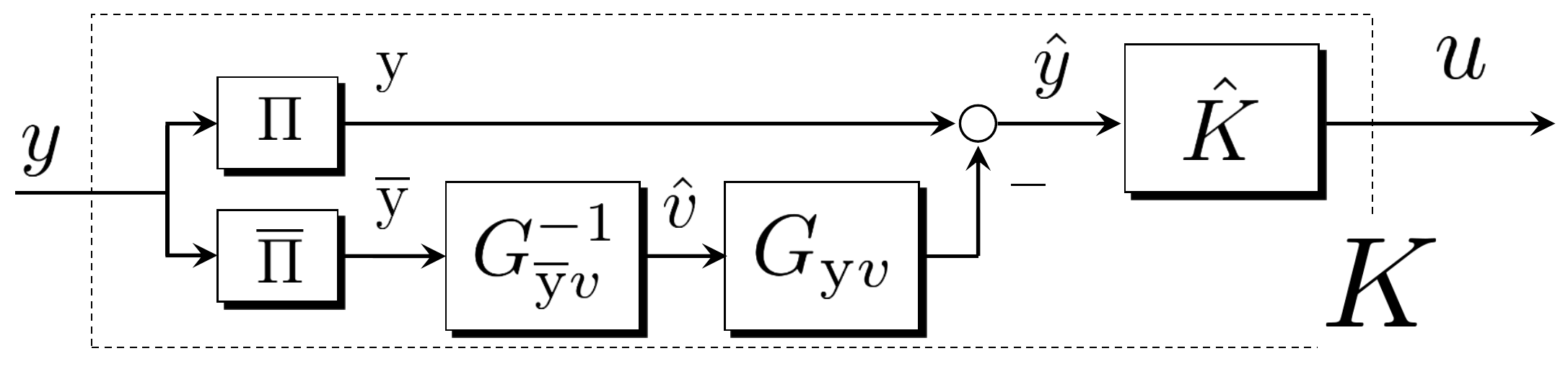}
\caption{Block diagram of output-feedback output-rectifying retrofit controllers with the proposed structure.
}
\label{fig:retro_str_of}
\end{figure}

Let $v\in\mathbb{R}^m$ and $y\in\mathbb{R}^p$.
Since it suffices to use $m$ independent outputs for reproducing $v$,
we take ${\rm y}:=\Pi y$ and $\ol{\rm y}:=\ol{\Pi}y$ with $\Pi\in \mathbb{R}^{(p-m)\times p}$ and $\ol{\Pi}\in \mathbb{R}^{m \times p}$, the latter of which is used for making an inverse system.
Take right-invertible matrices $\Pi$ and $\ol{\Pi}$ and their right inverses $\Pi\dg$ and $\ol{\Pi}\dg$ such that 
\begin{description}
\item[\textbf{C1}:] $\Pi\dg\Pi+\ol{\Pi}\dg\ol{\Pi}=I$,
\item[\textbf{C2}:] there exists an inverse system of $G_{\ol{\rm y}v}$ such that $G_{{\rm y}v}G_{\ol{\rm y}v}^{-1}$ is proper, where
\begin{equation}\label{eq:Gyv}
 G_{{\rm y}v}:=\Pi G_{yv},\quad G_{\ol{\rm y}v}:=\ol{\Pi} G_{yv},
\end{equation}
\end{description}
the existence of which will be shown below (see Lemma~\ref{lem:Pi}).
With those matrices, we replicate the interaction signal $v$ through $\hat{v}:=G_{\ol{\rm y}v}^{-1}\ol{\rm y}.$
Then the rectified output is given by
\begin{equation}\label{eq:int_meas}
 \hat{y} = {\rm y}-G_{{\rm y}v} \hat{v} = Ry,
\end{equation}
with
\begin{equation}\label{eq:R}
 R:=\Pi-G_{{\rm y}v}G_{\ol{\rm y}v}^{-1}\ol{\Pi},
\end{equation}
which specifies the controller structure as $K=\hat{K}R$.
The block diagram of the structured controller is illustrated by Fig.~\ref{fig:retro_str_of}, which is a replacement of Fig.~\ref{fig:retro_str}.
The objective of this section is to prove that all output-rectifying retrofit controllers are parameterized by the internal controller $\hat{K}$ with the proposed structure.

\subsection{Characterization of Controller Structure}

In this subsection, we characterize the structure of all output-feedback output-rectifying retrofit controllers.
We first show the existence of matrices that satisfy \textbf{C4} and \textbf{C5}.
The following lemma holds.
\begin{lem}\label{lem:Pi}
Let Assumptions~\ref{assum:sta} and~\ref{assum:inv} hold.
Then there exist matrices such that \textbf{C1} and \textbf{C2} hold.
\end{lem}
\begin{proof}
Let $g_i$ denote the $i$th row vector of $G_{yv}$.
From Assumption~\ref{assum:inv}, there exist $m$ row vectors of $G_{yv}$ that are linearly independent in the row vector space $\mathcal{R}^m$ over the field $\mathcal{R}$.
Denoting the index set of the $m$ rows by $\mathcal{I}\subset\{1,\ldots,p\}$,
we have
\[
 {\rm col}(g_i)_{i\notin \mathcal{I}} = H{\rm col}(g_i)_{i\in \mathcal{I}}
\]
with a possibly improper transfer matrix $H\in\mathcal{R}^{(p-m)\times m}$.
Let $\ol{\Pi}\in \mathbb{R}^{m\times m}$ and $\Pi \in \mathbb{R}^{(p-m)\times m}$ be the matrices to extract the $m$ rows and the other rows, respectively.
Then we have $H=G_{{\rm y}v}G_{\ol{\rm y}v}^{-1}$.
Thus, it suffices to find $m$ linearly independent row vectors with which $H$ becomes proper.

We demonstrate the procedure of choosing appropriate $m$ row vectors through an example for $m=2$ and $p=4$, which can easily be generalized to any case.
Denote the $(i,j)$th component of $H$ by $h_{ij}$ for $1\leq i \leq p-m$ and $1\leq j\leq m$.
Supposing that $g_1$ and $g_2$ are linearly independent, we have
\begin{equation}\label{eq:g412}
 g_3 = h_{11}g_1+h_{12}g_2,\quad g_4 = h_{21}g_1+h_{22}g_2.
\end{equation}
Denote the relative degree of $h_{ij}$ by $r_{ij}$, which can be a negative integer.
Let us focus on the equation on $g_3$ and take $\ol{r}_1:=\min_{1\leq j\leq m}r_{1j}$.
We suppose $\ol{r}_1=r_{11}$ in this example.
If $\ol{r}_1\geq 0$, we proceed to the next row.
If $\ol{r}_1<0$, by dividing the equation by $h_{11}$, we have
\begin{equation}\label{eq:g132}
g_1=h_{11}^{-1}g_3-h_{11}^{-1}h_{12}g_2,
\end{equation}
where the coefficients are proper.
In this case, $g_2$ and $g_3$ are linearly independent because $h_{11}\neq 0$. 
Substituting~\eqref{eq:g132} into~\eqref{eq:g412} yields
\[
 g_1=h_{11}'g_3+h_{12}'g_2,\quad g_4= h_{21}'g_3+h_{22}'g_2
\]
where
\begin{equation}\label{eq:hs}
 \begin{array}{l}
 h_{11}':=h_{11}^{-1},\quad h_{12}':=-h_{11}^{-1}h_{12},\\
 h_{21}':=h_{21}h_{11}^{-1},\quad h_{22}':=h_{12}-h_{21}h_{11}^{-1}h_{12}.
 \end{array}
\end{equation}
Similarly, denote the relative degrees of the coefficients by $r'_{ij}$, take $\ol{r}'_2:= \min_{1\leq j \leq m}r'_{2j}$, and suppose $\ol{r}'_2=r'_{22}$.
If $\ol{r}'_2\geq 0$, the coefficients in~\eqref{eq:hs} are proper and hence $g_2$ and $g_3$ satisfy the requirement.
If $\ol{r}'_2< 0$, we obtain
\[
 g_1=h_{11}''g_3+h_{12}''g_4,\quad g_2= h_{21}''g_3+h_{22}''g_4
\]
through the same procedure, where the coefficients are proper.
Because $h_{22}'\neq 0,$ $g_3$ and $g_4$ are linearly independent.
Therefore, $g_3$ and $g_4$ satisfy the requirement.
\end{proof}

Next, we characterize all solutions to the linear equation~\eqref{eq:QG} without the condition $Q\in \mathcal{RH}_{\infty}$.
\begin{lem}\label{lem:of_str}
Let Assumptions~\ref{assum:sta} and~\ref{assum:inv} hold.
The Youla parameter $Q$, which is possibly unstable, satisfies~\eqref{eq:QG} if and only if there exists $\hat{K}\in\mathcal{RP}$ such that $K=\hat{K}R$ where $R$ is defined in~\eqref{eq:R}.
\end{lem}
\begin{proof}
We seek for all solutions to $KG_{yv}=0.$
We have
\[
\begin{array}{cl}
 RG_{yv}\hs=\Pi(I-G_{yv}G_{\ol{\rm y}v}^{-1}\ol{\Pi})G_{yv}\\
 \hs = \Pi G_{yv}-\Pi G_{yv}G_{\ol{\rm y}v}^{-1}G_{\ol{\rm y}v}\\
 \hs =0,
\end{array}
\]
and hence the sufficiency holds.
For the necessity, consider
\[
 S:=\left[
 \begin{array}{c}
 R\\
 \ol{\Pi} 
 \end{array}
 \right],\quad S^{-1}:=\left[\Pi\dg\ G_{yv}G_{\ol{\rm y}v}^{-1} \right],
\]
each of which is the inverse of the other because
\[
\begin{array}{cl}
 S^{-1}S \hs = \Pi\dg\Pi - \Pi\dg G_{{\rm y}v}G_{\ol{\rm y}v}^{-1} \ol{\Pi}+G_{yv}G_{\ol{\rm y}v}^{-1}\ol{\Pi}\\
  \hs = \Pi\dg\Pi + (I-\Pi\dg\Pi)G_{yv}G_{\ol{\rm y}v}^{-1}\ol{\Pi}\\
  \hs = \Pi\dg\Pi + \ol{\Pi}\dg\ol{\Pi}G_{yv}G_{\ol{\rm y}v}^{-1}\ol{\Pi}\\
  \hs = \Pi\dg\Pi + \ol{\Pi}\dg G_{\ol{\rm y}v}G_{\ol{\rm y}v}^{-1} \ol{\Pi}\\
  \hs = \Pi\dg\Pi + \ol{\Pi}\dg \ol{\Pi}\\
  \hs = I
\end{array}
\]
from \textbf{C1}.
Since $S$ and $S^{-1}$ are proper from \textbf{C2}, $S$ is unimodular, i.e., invertible in the ring of $\mathcal{RP}$.
Because $S$ is unimodular, for any $K\in\mathcal{RP}$, there always exists $\tilde{K}=\left[\hat{K}\ \check{K}\right]\in\mathcal{RP}$ such that $K=\tilde{K}S=\hat{K}R+\check{K}\ol{\Pi}$.
It suffices to show that $\check{K}=0$.
From $KG_{yv}=0$ and $RG_{yv}=0$, the latter of which has been proven in the sufficiency part, it turns out that $\check{K}G_{\ol{\rm y}v}=0.$
Since $G_{\ol{\rm y}v}$ is invertible from \textbf{C2}, this equation is equivalent to $\check{K}=0$, which proves the claim.
\end{proof}
Lemma~\ref{lem:of_str} implies that all output-feedback output-rectifying retrofit controllers have the structure depicted in Fig.~\ref{fig:retro_str_of}.

\subsection{Identification of Class of ${\it \hat{K}}$: Minimum-phase ${\it G_{{\rm y}v}}$}

What remains to do is to identify the class of the parameter $\hat{K},$ which is relevant to the condition $Q\in\mathcal{RH}_{\infty}$.
Define
\[
 \hat{Q}:=(I-\hat{K}G_{{\rm y}u})^{-1}\hat{K},
\]
the Youla parameter of the internal controller $\hat{K}$ rather than the overall controller $K$,
where
\begin{equation}\label{eq:Gyu}
 G_{{\rm y}u}:=RG_{yu}.
\end{equation}
The following lemma reduces the condition on $Q$ to that on $\hat{Q}$.
\begin{lem}\label{lem:QQGG}
Let $K=\hat{K}R$.
The condition $Q\in\mathcal{RH}_{\infty}$ holds if and only if $\hat{Q}\in\mathcal{RH}_{\infty}$
and
\begin{equation}\label{eq:QQGG}
 \hat{Q}G_{{\rm y}v}G_{\ol{\rm y}v}^{-1}\in\mathcal{RH}_{\infty}.
\end{equation}
\end{lem}
\begin{proof}
From the definition, we have
\[
 \begin{array}{cl}
 Q \hs = (I-KG_{yu})^{-1}K\\
  \hs = (I-\hat{K}RG_{yu})^{-1}\hat{K}R\\
  \hs = \hat{Q}\Pi-\hat{Q}G_{{\rm y}v}G_{\ol{\rm y}v}^{-1}\ol{\Pi}.
 \end{array}
\]
Hence, the sufficiency is obvious.
From the equation, it turns out that $\hat{Q}=Q\Pi\dg$ and $\hat{Q}G_{yv}G_{\ol{\rm y}v}^{-1} = Q\ol{\Pi}\dg$ since $\Pi\ol{\Pi}\dg=0$ and $\ol{\Pi}\Pi\dg=0$.
Thus the necessity holds.
\end{proof}
From Assumption~\ref{assum:sta}, $G_{{\rm y}v}$ is stable.
Hence if $G_{\ol{\rm y}v}$ is minimum phase, the latter condition~\eqref{eq:QQGG} can be removed, and a simple characterization can be obtained.
\begin{theorem}\label{thm:main1}
Let Assumptions~\ref{assum:sta} and~\ref{assum:inv} hold.
Assume also that $G_{\ol{\rm y}v}$ is minimum phase, i.e., it contains no unstable zeros.
Then the controller $K$ is an output-rectifying retrofit controller if and only if there exists $\hat{K}\in\mathcal{S}(G_{{\rm y}u})$ such that $K=\hat{K}R$ where $R$ is defined in~\eqref{eq:R}.
\end{theorem}
\begin{proof}
From the assumption, $R$ in~\eqref{eq:R} is stable.
Thus $\hat{Q} \in \mathcal{RH}_{\infty}$ is equivalent to $\hat{K}\in\mathcal{S}(G_{{\rm y}u})$.
Therefore, Lemma~\ref{lem:of_str} and Lemma~\ref{lem:QQGG} lead to the claim.
\end{proof}

\subsection{Identification of Class of $\hat{K}$: Non-minimum phase $G_{{\rm y}v}$}

In contrast to the previous case, when $G_{{\rm y}v}$ is non-minimum phase, which may result in unstable $G_{{\rm y}v}G_{\ol{\rm y}v}^{-1}$, the condition~\eqref{eq:QQGG} cannot be simplified in a straightforward manner.
Thus the same characterization of the class of $\hat{K}$ is unavailable.

Because it is difficult to further investigate the structure only by frequency domain analysis,
we derive the normal form~\cite{Mueller2009Normal} of $RG_{yu}$ and $G_{{\rm y}v}G_{\ol{\rm y}v}^{-1}$ in the time domain.
As a preliminary step, we introduce the notion of relative degree of multi-input and multi-output systems.
\begin{defin}\label{def:rel}
Consider a strictly proper transfer matrix $G\in \mathcal{RP}^{m\times m}$
with a realization $(A,B,C)$.
Then $(r_1,\ldots,r_m)$ is said to be the relative degree of $G$ if for $i=1,\ldots,m$,
\[
 c_iA^kB=0,\quad c_iA^{r_i-1}B \neq 0,\quad \forall k \leq r_i-2 
\]
and ${\rm col}(c_iA^{r_i-1}B)_{i=1}^m$ is nonsingular where $c_i$ stands for the $i$th row vector of $C$.
\end{defin}
Let $(r_1,\ldots,r_m)$ be the relative degree of $G_{\ol{\rm y}v}$, where $r_i\geq 1$ for any $i=1,\ldots,m$.
Under this assumption, we consider a coordinate transformation for the state-space representation of $G_{{\rm y} v}$ and $G_{\ol{\rm y}v}$ in~\eqref{eq:Gyv}.
Take the matrices
\[
 T:={\rm col}(T_i)_{i=1}^m,\quad T_i:= {\rm col}(e_i\ol{\Pi}CA^{j-1})_{j=1}^{r_i}
\]
with the $m$-dimensional $i$th canonical row vector $e_i$.
It can be shown that there exists $\ol{T}$ such that $T$ and $\ol{T}$ complete the coordinates, $\ol{T}L=0$, and $\ol{\Pi}C\ol{T}\dg=0$~\cite{Mueller2009Normal}.
Consider the coordinate transformation ${\rm x}\mapsto(z,\xi)$ with $\xi:=T{\rm x}$ and $z:=\ol{T}{\rm x},$ we have a particular realization
\begin{equation}\label{eq:normal}
\arraycolsep=3pt
\left\{
 \begin{array}{cl}
 \left[
 \begin{array}{c}
 \dot{\xi}\\
 \dot{z}
 \end{array}
 \right] \hs= \left[
 \begin{array}{cc}
 A_{\xi \xi} & A_{\xi z}\\
 A_{z \xi} & A_{zz}
 \end{array}
 \right]
 \left[
 \begin{array}{c}
 \xi\\
 z
 \end{array}
 \right]+
 \left[
 \begin{array}{c}
 TL\\
 0
 \end{array}
 \right]v+
 \left[
 \begin{array}{c}
 TB\\
B_{z u}
 \end{array}
 \right]u\\
 {\rm y} \hs = C_{{\rm y}\xi} \xi + C_{{\rm y}z} z,\quad
 \ol{\rm y} = \ol{\Pi}C T\dg \xi
 \end{array}
 \right.
\end{equation}
with
\[
\begin{array}{lll}
A_{\xi \xi}:= TAT\dg,& A_{\xi z}:= TA\ol{T}\dg,& A_{z \xi}:= \ol{T}AT\dg,\\
A_{zz}:= \ol{T}A\ol{T}\dg, & B_{z u}:=\ol{T}B, & C_{{\rm y}\xi}:=\Pi C T\dg,\\
C_{{\rm y}z}:=\Pi C \ol{T}\dg. & &
\end{array}
\]
The realization~\eqref{eq:normal} has an advantage that $\xi$ can be represented as a simple derivative of $\ol{\rm y}$ and $u$.
Indeed, we have
\begin{equation}\label{eq:xi}
 \xi=\mathcal{D}_{\ol{\rm y}}\ol{\rm y}-\mathcal{D}_uu
\end{equation}
where the differential operator $\mathcal{D}_{\ol{\rm y}}$ is defined by
\[
 \begin{array}{l}
 \mathcal{D}_{\ol{\rm y}}:={\rm diag}(\mathcal{D}_i)_{i=1}^m,\quad \mathcal{D}_i:= {\rm col}(d^{j-1}/{dt}^{j-1})_{j=1}^{r_i},\\
 \end{array}
\]
and $\mathcal{D}_u$ is defined in a similar manner to be compatible with~\eqref{eq:Gss}.

\begin{figure}[t]
\centering
\includegraphics[width = .95\linewidth]{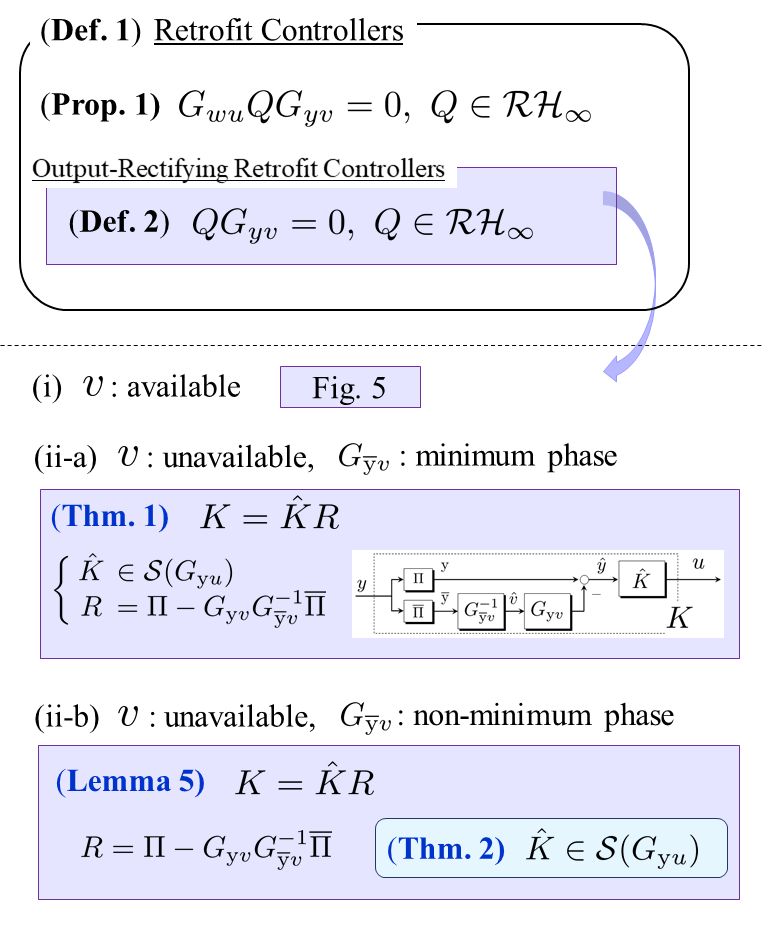}
\caption{
Refined Venn diagrams of retrofit controllers based on the obtained results.
}
\label{fig:Venn2}
\end{figure}

Based on the above preparation, we can obtain compact representations of $G_{{\rm y}u}$ and $G_{{\rm y}v}G_{\ol{\rm y}v}^{-1}$ in the time domain.
The following lemma holds.
\begin{lem}\label{lem:td_representation}
The system $G_{{\rm y}u}:u\mapsto \hat{y}$ can be represented by
\begin{equation}\label{eq:RGyu}
  G_{{\rm y}u}:\left\{
 \begin{array}{cl}
 \dot{\phi}\hs = A_{\rm zz} \phi+(B_{zu}+\mathcal{D}_u)u\\
 \hat{y}\hs =C_{{\rm y}z} \phi - C_{{\rm y}\xi} \mathcal{D}_uu
 \end{array}
 \right.
\end{equation}
and the dynamics of $G_{{\rm y}v}G_{\ol{\rm y}v}^{-1}:\ol{\rm y}\mapsto \hat{{\rm y}}$ can be represented by
\begin{equation}\label{eq:ss_GGinv}
G_{{\rm y}v}G_{\ol{\rm y}v}^{-1}:\left\{
 \begin{array}{cl}
 \dot{\zeta}\hs = A_{\rm zz} \zeta+A_{{\rm z}\xi} \mathcal{D}_{\ol{\rm y}}\ol{\rm y}\\
 \hat{{\rm y}}\hs = C_{{\rm y}z}\zeta + C_{{\rm y}\xi} \mathcal{D}_{\ol{\rm y}}\ol{\rm y}.
 \end{array}
 \right.
\end{equation}
\end{lem}
\begin{proof}
We first suppose $u=0$.
Then substituting $\xi=\mathcal{D}_{\ol{\rm y}}\ol{\rm y}$ to~\eqref{eq:normal} yields~\eqref{eq:ss_GGinv}.
Next, suppose $v=0$.
Then
\[
 \hat{y}=-C_{{\rm y}z} \zeta+{\rm y}-C_{{\rm y}\xi} \mathcal{D}_{\ol{\rm y}}\ol{\rm y}
\]
where $z,{\rm y},\ol{\rm y}$ obey the dynamics~\eqref{eq:normal} and~\eqref{eq:ss_GGinv}.
We have
\[
 \begin{array}{cl}
 \hat{y} \hs = C_{{\rm y}z}(z-\zeta)+C_{{\rm y}\xi} \xi - C_{{\rm y}\xi} \mathcal{D}_{\ol{\rm y}}\ol{\rm y}\\
 \hs = C_{{\rm y}z}(z-\zeta)+C_{{\rm y}\xi}(\mathcal{D}_{\ol{\rm y}}\ol{\rm y}-\mathcal{D}_uu)- C_{{\rm y}\xi} \mathcal{D}_{\ol{\rm y}}\ol{\rm y}\\
 \hs = C_{{\rm y}z}(z-\zeta) - C_{{\rm y}\xi}\mathcal{D}_u u
 \end{array}
\]
from~\eqref{eq:xi}.
Define $\phi:=z-\zeta$, whose dynamics is given by
\[
 \begin{array}{cl}
 \dot{\phi} \hs = A_{\rm zz}\phi+A_{{\rm z}\xi}\xi + B_{zu}u - A_{{\rm z}\xi} \mathcal{D} \ol{\rm y}\\
 \hs =A_{\rm zz}\phi+(B_{zu}+\mathcal{D}_u)u,
 \end{array}
\]
which leads to~\eqref{eq:RGyu}.
\end{proof}

What should be emphasized in Lemma~\ref{lem:td_representation} is that $RG_{yu}$ and $G_{{\rm y}v}G_{{\ol{\rm y}}v}^{-1}$ share the state matrix, or ``$A$-matrix,'' which is given as a reduced matrix of the original matrix $A$ through the projection.
Thus, the following lemma holds.
\begin{lem}\label{lem:of_Kclass}
Let Assumptions~\ref{assum:sta} and~\ref{assum:inv} hold and the controller be given by $K=\hat{K}R$.
If $\hat{K}$ stabilizes~\eqref{eq:RGyu}, then the condition in~\eqref{eq:QQGG} holds.
\end{lem}
\begin{proof}
Assume that $\hat{K}:\hat{y}\mapsto u$ stabilizes~\eqref{eq:RGyu}.
Then from Lemma~\ref{lem:td_representation}, the state matrix of $\hat{Q}$ and $\hat{Q}G_{{\rm y}v}G_{{\ol{\rm y}}v}^{-1}$ with the representation~\eqref{eq:RGyu} and~\eqref{eq:ss_GGinv} becomes stable.
Thus, both $\hat{Q}$ and $\hat{Q}G_{{\rm y}v}G_{{\ol{\rm y}}v}^{-1}$ belong to $\mathcal{RH}_{\infty}$.
\end{proof}
Note that, although the representations~\eqref{eq:RGyu} and~\eqref{eq:ss_GGinv} contain differential operators, the transfer matrices are guaranteed to be proper from the frequency-domain representations.
Note also that the state matrix can be invariant even if we represent them without the differential operators.

From Lemmas~\ref{lem:of_str} and~\ref{lem:of_Kclass}, the following theorem holds.
\begin{theorem}\label{thm:gen}
Let Assumptions~\ref{assum:sta} and~\ref{assum:inv} hold.
The controller $K$ is an output-feedback output-rectifying retrofit controller if there exists $\hat{K}$ such that $K=\hat{K}R$ and $\hat{K}$ is a stabilizing controller for~\eqref{eq:RGyu}.
\end{theorem}
In Theorem~\ref{thm:gen}, only a sufficient condition is provided because the stabilizing condition on $\hat{K}$ is slightly stricter than $\hat{Q}\in \mathcal{RH}_{\infty}$ and $\hat{Q}G_{{\rm y}v}G_{\ol{\rm y}v}^{-1}\in \mathcal{RH}_{\infty}$.
When $G_{{\rm y}u}$ is unstable, the condition means stability of the corresponding four transfer matrices, including $\hat{Q}$, which leads to $\hat{Q}\in \mathcal{RH}_{\infty}$ and $\hat{Q}G_{{\rm y}v}G_{\ol{\rm y}v}^{-1}\in \mathcal{RH}_{\infty}$.
The gap is caused by the possible instability of $G_{\ol{\rm y}v}^{-1}$ as claimed in the beginning of this subsection.

\subsection{Summary on Parameterization}
The obtained results are summarized in Fig.~\ref{fig:Venn2}.
This figure indicates that the controller structure composed of an internal controller $\hat{K}$ and a rectifier $R$ is necessary and sufficient for the constraint~\eqref{eq:QG} in any case.
The other condition $Q\in\mathcal{RH}_{\infty}$ is equivalent to the stabilizing capability of $\hat{K}$ in most cases, although the condition becomes only sufficient when $G_{\ol{\rm y}v}$ is non-minimum phase.
Finally, the rectifier $R$ can systematically be constructed from $G_{yv}$, and the system stabilized by $\hat{K}$ has a realization as a reduced-order model of $G_{yu}$ in any case.
This parameterization derives an output-rectifying retrofit controller synthesis algorithm, described in Algorithm~1.

\begin{algorithm}[th]
\caption{Retrofit Controller Synthesis Algorithm}
\begin{algorithmic}[1]
\REQUIRE{$G$: subsystem of interest}
\ENSURE{$K$: output-rectifying retrofit controller}
\STATE Find $\Pi$ and $\Pi$ with \textbf{C1} and \textbf{C2} based on Lemma~\ref{lem:Pi}
\STATE Build $G_{{\rm y}v},G_{\ol{\rm y}v},G_{{\rm y}u}$ defined in~\eqref{eq:Gyv} and~\eqref{eq:Gyu}
\STATE Choose an internal controller $\hat{K}$ in $\mathcal{S}(G_{{\rm y}u})$
\STATE Construct the controller $K=\hat{K}R$ with $R$ defined in~\eqref{eq:R}
\end{algorithmic}
\end{algorithm}

\section{Numerical Example}
\label{sec:num}

We consider a network system where each component's dynamics is given as a second-order system.
Second-order network systems are, for instance, used as a simple model of power transmission systems~\cite{Kundur94}.
Let $N=50$ be the number of the components.
We suppose that each component $\Sigma_k$ for $k=1,\ldots,N$ is represented by
\[
 \Sigma_k: m_k\ddot{\theta}_k+d_k\dot{\theta}_k+v_k+u_k=0,\quad y_k=\omega_k
\]
where $\theta_k\in\mathbb{R}$ and $\omega_k:=\dot{\theta}_k\in\mathbb{R}$ are the state, $v_k\in\mathbb{R}$ is an interaction signal given by
\[
 v_k=\sum_{l\in\mathcal{N}_k}\alpha_{k,l}(\theta_k-\theta_l)
\]
with the $k$th subsystem's neighbourhood $\mathcal{N}_k$,
$u_k\in\mathbb{R}$ and $y_k\in\mathbb{R}$ are the control input and the measurement output, respectively.
The parameters are specifically given by
\[
 m_k=1,\quad d_k=0.5,\quad \alpha_{k,l}=1
\]
for any $k,l=1,\ldots,N$.
The control objective is to attenuate effects from disturbance to $\omega_k$.
Let the first half components be the subsystem of interest in retrofit control, namely, $G$ in Fig.~\ref{fig:pre_sys}, and the others be the environment $\ol{G}$.

\begin{figure}[t]
\centering
\subfloat[][The graph at the beginning with the smallest-dimensional interaction signal betweeen the subsystem of interest and the environment.]{\includegraphics[width=.98\linewidth]{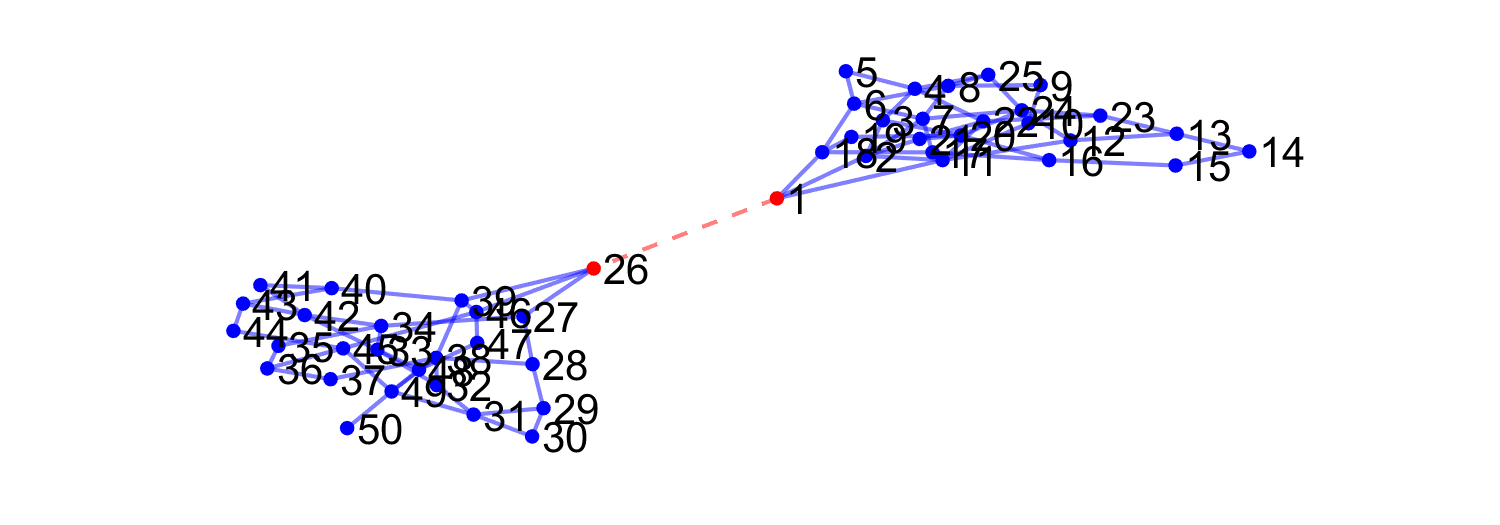}\label{subfig:graph_1}}
\\
\subfloat[][The graph with the largest-dimensional interaction signal.]
{\includegraphics[width=.98\linewidth]{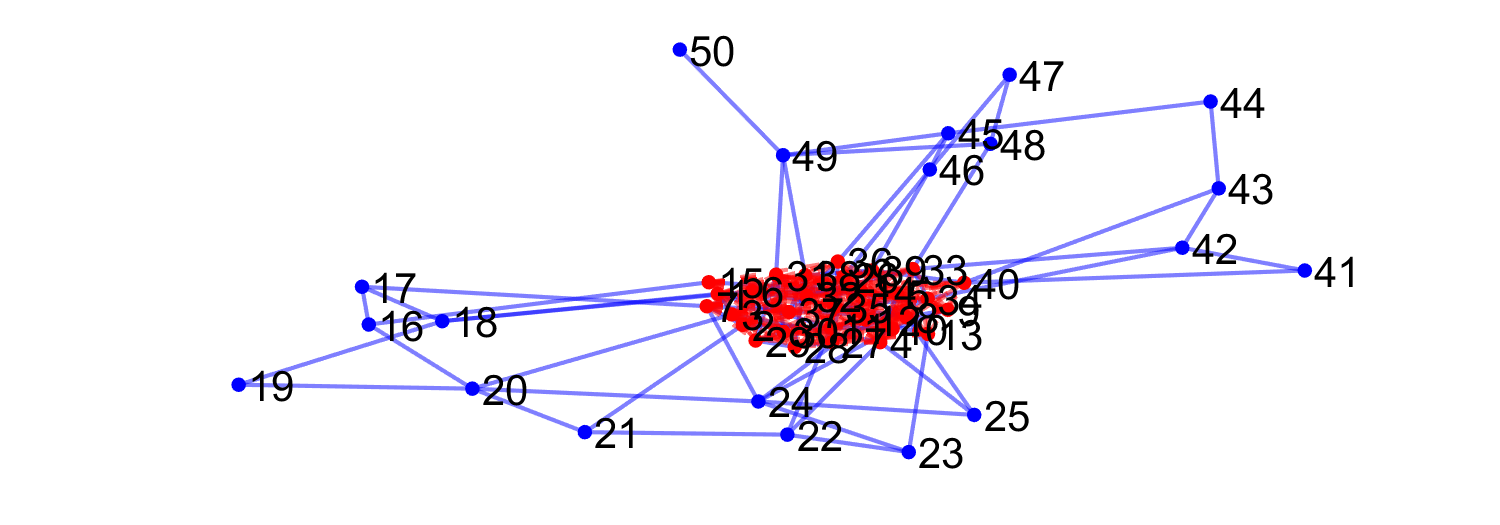}\label{subfig:graph_2}}
\caption[]{The beginning and final graphs to be considered.
The graph at the beginning has the smallest-dimensional interaction signal while the final graph has the largest-dimensional interaction signal.
The broken lines depict the interaction between the subsystem of interest $G$ and the environment $\ol{G}$.
}
\label{fig:graphs}
\end{figure}

We confirm the effectiveness of retrofit control and also compare performances achieved using the newly developed retrofit controllers, namely, retrofit controllers without interaction measurement, to the existing ones with interaction measurement.
The internal controller is designed to be a linear quadratic regulator with a state observer under certain weights.
To highlight the impacts of measurement of the interaction signal, we consider multiple networks where the dimension of the interaction signal $v$ increases gradually.
The graph at the beginning is illustrated in Fig.~\ref{subfig:graph_1} where $G$ and $\ol{G}$ are interconnected through a single edge between two components.
Subsequently, adding an edge between the subsystems we obtain a new graph where the dimension of the interaction signal is larger than the previous one.
By repeating this process, we obtain the final graph illustrated in Fig.~\ref{subfig:graph_2} where the first fifteen components of $G$ and $\ol{G}$ are interconnected to each other.
We consider the fifteen graphs for performance evaluation.

\begin{figure}[t]
\centering
\includegraphics[width = .95\linewidth]{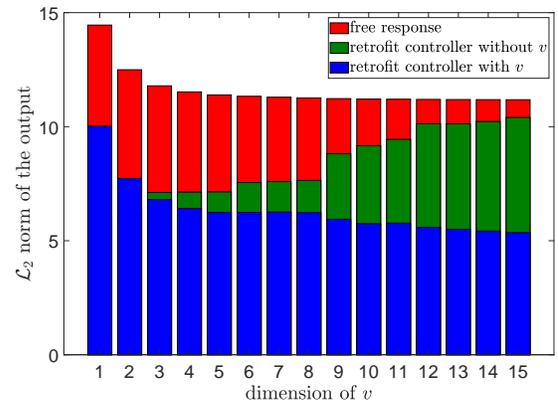}
\caption{
The control performances in the cases without controllers, with a newly developed retrofit controller without interaction measurement, and with an existing retrofit controller with interaction measurement for the fifteen graphs.
}
\label{graph:L2s}
\end{figure}

The control performances are compared in Fig.~\ref{graph:L2s}, where the horizontal and vertical axes correspond to the dimension of $v$ and the $\mathcal{L}_2$ norm of $(\omega_1,\ldots,\omega_N)$ against an initial disturbance, respectively.
The red, green, and blue bars are overlapped and correspond to the cases without controllers, with the retrofit controller without interaction measurement, and the retrofit controller with interaction measurement, respectively.
Observations available from Fig.~\ref{graph:L2s} are the following twofold.
First, when the dimension of the interaction signal is not very large, the control performance is significantly improved through both of the retrofit controllers.
This result indicates practical impacts of retrofit control even without interaction measurement.
The other one is that, although the performance of the retrofit controller without interaction measurement is almost the same as that of the one with interaction measurement in the small interaction case, the performance is gradually worse as edges between $G$ and $\ol{G}$ are added.
This performance deterioration is caused because the dimension of the rectified output $\hat{y}$ in~\eqref{eq:int_meas} is reduced as the dimension of $v$ increases.

\section{Conclusion}
\label{sec:conc}

This study has investigated a parameterization of all output-rectifying retrofit controllers in the general output-feedback case based on system inversion.
The derived results compliment the existing findings in the sense that all output-rectifying retrofit controller can be parameterized with an internal controller and designed through existing controller synthesis methods.
A possible future work includes developing a numerical synthesis method of general retrofit controllers, for which recent results on controller parameterization~\cite{Wang2019A,Furieri2019An,Zheng20On} would be helpful.

\bibliography{sshrrefs}
\bibliographystyle{IEEEtran}

\end{document}